\documentclass[%
 aip,
amsmath,amssymb,
]{revtex4-2}

\usepackage{graphicx}
\usepackage{color}
\usepackage{dcolumn}
\usepackage{bm}

\usepackage[utf8]{inputenc}
\usepackage[T1]{fontenc}
\usepackage{mathptmx}
\usepackage{appendix}

\begin{document}


\title[Paschen Curve for $\mathrm{H_2}$ and $\mathrm{D_2}$]{Modeling  the low-pressure high-voltage branch of the Paschen curve for hydrogen and deuterium}
\author{A.~V.~Khrabrov}
\email{skhrab@gmail.com}
\affiliation{Princeton Plasma Physics Laboratory, Princeton NJ 08543 USA}

\author{D.~J.~Smith}
\affiliation{GE Vernova Advanced Research, Niskayuna NY 12309 USA}

\author{I.~D.~Kaganovich}
\affiliation{Princeton Plasma Physics Laboratory, Princeton NJ 08543 USA}
%



\date{\today}

\begin{abstract}
A physical and numerical model of the Townsend discharge in molecular hydrogen and deuterium has been developed to meet the needs of designing a plasma-based switching device for power grid application. The model allows to predict the low-pressure branch of the Paschen curve for applied voltage in the range of several hundred kiloVolts.
In the regime of interest, electrons are in a runaway state and ionization by ions and fast neutrals sustains the discharge. It was essential to correctly account for both gas-phase and surface interactions (electron emission and electron back-scattering), especially in terms of their dependence on particle energy. The model yields results consistent with prior data obtained for lower voltage. The three-species (electrons, ions and fast neutrals) model successfully captures the essential physics of the process. 
\end{abstract}

\maketitle

%


\section{\label{sec:Intro} Introduction}

 The electric transmission and distribution grid needs to be substantially expanded and modernized to accommodate the shift towards electrification and the integration of renewable energy power sources. At present, the electric grid is primarily powered by alternating current (AC) electricity. However, direct current (DC) transmission and distribution of electric power has advantages over AC technology, including lower distribution losses, higher capacity, and improved stability, as well as compatibility with renewable sources, and industrial applications. The primary obstacle to the wider implementation of DC transmission is the lack of suitable medium voltage (MV) and high voltage (HV) circuit breakers to isolate faults and allow the remainder of the system to continue to operate.
 
Commercially viable solutions for DC circuit breakers must be reliable, low loss, high power density, and affordable. Unlike AC systems which can utilize the zero-crossing current that occurs twice in every cycle, the fault current increases rapidly (~kA/ms) in DC systems, limited only by the total energy stored in the system impedance. A DC circuit breaker must provide an opposing voltage to create a current-zero and fast operation is essential to minimize the peak current. There are several different types of MV and HV DC circuit breakers under development \cite{CIGRE2020, CIGRE2023, CIGRE2022}; we have investigated the use of a gas discharge tube (GDT) device [4] as a potential option. The GDT is a hermetic assembly of metallic electrodes within a ceramic insulator filled with a specified gas to enable transition between an insulating state and a conducting plasma \cite{Goebel1996}.

For GDT standoff, the separation between the anode and designated cathode, $d$, must be large enough to avoid vacuum breakdown (typical onset at $80-120 \mathrm {kV/cm}$). We may select the appropriate electrode separation based on the required device voltage rating. Importantly, the device operates on the left side of the Paschen curve, setting an upper value of $pd$. We may therefore define the upper limit on pressure, $p$, for a given GDT design. The conductive properties of the gas are also of great importance to ensure high plasma conduction current for the device. Here, we have considered hydrogen and deuterium as candidate fill gases for the GDT. 

A predictive model for holding off the high voltage between the anode and the control grid needs to be newly developed for the given operating conditions, namely $100 - 300~\mathrm{kV}$ under open circuit. The general approach and the numerical code employed in this study were previously applied to investigate the low-pressure breakdown in helium \cite{Liang2017} and also in argon \cite{Jin2022}. In the low-pressure regime of interest, such model needs to be based directly on the kinetic description of the particle species involved in the process. This is because particle velocity distributions are strongly anisotropic and, for charged species, and non-local, i.e., also not completely determined by the local electric field. The electrons, in particular, are in a runaway regime and can attain much higher energy than the value at the maximum of ionization cross section ($\sim$50 eV). Therefore, their contribution to gas-phase ionization is primarily due to electrons  inelastically reflected and confined in the vicinity of the anode. The ionization cascade in the volume is sustained through charge exchange formation of fast neutrals, and by impact ionization by ions and fast neutrals. For ions, the mean free path is still short compared to the discharge gap, but a quasi-equilibrium distribution forms only within several free-path lengths from the anode. Same applies to the fast neutral species as their velocity distribution is determined by that of the ions. A fully kinetic model is still needed to describe this behavior. 

By seeking a steady-state of the discharge for a given voltage, one can find the $pd$ value corresponding to the breakdown threshold, i.e. the corresponding point on the Paschen curve. In this report we present the resulting prediction in the range of 40 to 300 kV for hydrogen and 80 to 600 kV for deuterium. The limits chosen for deuterium are based on the approximate mass scaling discussed in the Appendix and briefly in the main body of the article. 

Performance of the kinetic model depends on reliable input data for the underlying interactions in the gas phase and with material surfaces. We made an effort to seek and utilize the best available information. The specifics of the kinetic model for high-voltage Townsend discharge hydrogen are laid out in Sec.~\ref{sec:model}, followed by presenting the results in Sec.~\ref{sec:results}. The numerical framework, already presented in full detail in our prior work \cite{Liang2017,Jin2022}, is outlined when discussing the results in Sec.~\ref{sec:PIC}, which, in turn, is followed by presenting conclusions.

\section{\label{sec:model} Physical model and input data}
\subsection{General description}
\begin{figure*}[htbp]
\includegraphics[scale=0.4, angle = 0]{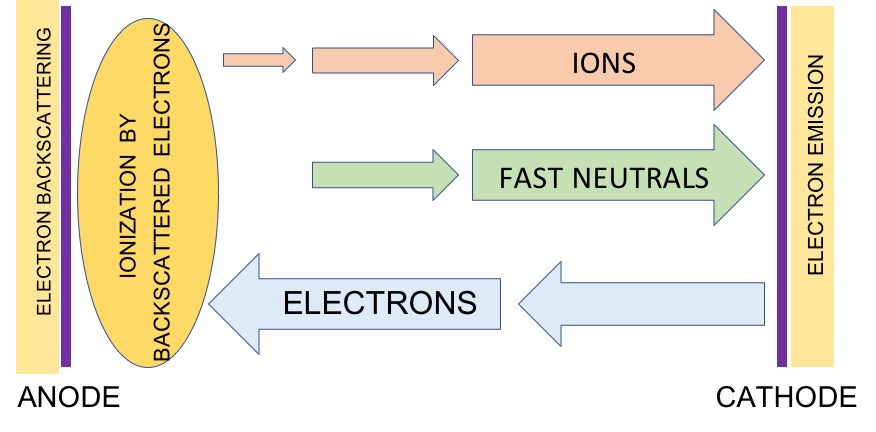}
\caption{Schematic of the high-voltage discharge. Arrows represent fluxes of the particle species. The increasing size represents  multiplication of the flux. For ions and fast neutrals, the flux at the anode surface is zero. The electron flux starting at the cathode is several times the impinging ion flux due to the high emission yield of ions and neutrals, and relative multiplication of the electron flux is not large.} 
\label{fig:schematic} 
\end{figure*}

The kinetic model is based on tracking the particles in a one-dimensional discharge gap \cite{Liang2017,Jin2022}. The discharge gap and key kinetic phenomena taking place in it are depicted schematically in Fig.~\ref{fig:schematic}.
The specific inter-electrode distance was set at 1.4 cm, same as in the study by Xu {\it et al.} \cite{Liang2017} of a cold-cathode magnetron switch utilizing helium as the working gas. At the present level of physical detail, the $pd$ scaling rule will apply. In the specific case of hydrogen gas, there is, in fact, a controversy in experimental literature on the subject of $pd$ scaling \cite{JETP57,McClure59,JETP62}. Addressing this issue would require a much more detailed kinetic description of the discharge, especially considering that the underlying mechanism responsible for breaking the scaling law has not yet been identified. Nevertheless, the model utilized in this study (just like similar models developed in the past) still has predictive value in terms of guiding the engineering development, because the resulting low-pressure branch of the Paschen curve is very steep, that is, confined within a narrow range of $pd$ values. An early example of experimental results illustrating this property for applied potentials up to to $2~\mathrm{kV}$ is the work performed by Armstrong and Bennett\cite{PPPL97}.

In our model, the anode is situated on the left at $x=0$ and the cathode, which represents the control grid of the gas switch, is on the right at $x=d$. The material of both electrodes is molybdenum.
The model accounts for volumetric ionization (1) by electrons, (2) by molecular ions, and (3) by fast neutral molecules. The surface interactions are (1) inelastic reflection of energetic electrons from the anode, and (2) ion- and (3) fast-neutral-induced secondary emission  from the cathode. All angular scattering of fast particles in gas-phase collisions is neglected, same as the prior study of the high-voltage breakdown in helium \cite{Liang2017} in which two of the authors were involved. Neglecting the scattering is of much less influence than that of the uncertainties in the input data. 

The primary mechanism of sustaining the discharge is the cascade of charge-exchange and ionization by heavy species in which an ion produces several (about 10) fast neutrals, each capable of ionizing thermal molecules of the background gas. The resulting ions (gaining energy from the electric field) and fast neutrals produce further ionization. 

\subsection{Gas-phase interactions}
\begin{figure*}[htbp]
\includegraphics[scale=0.6]{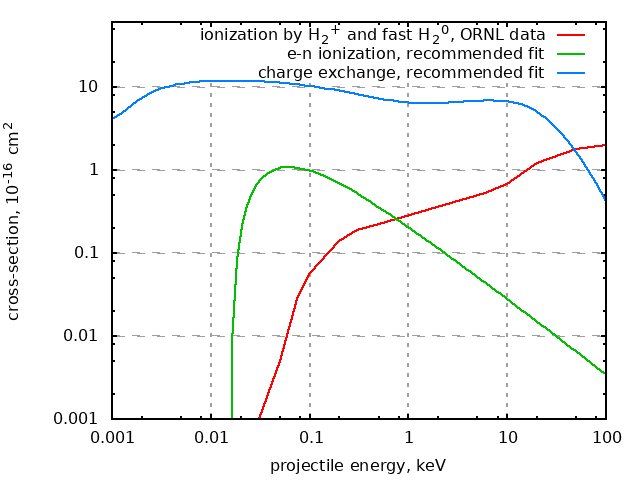}
\caption{Collision processes accounted for in the physical model. Analytical fitting functions are from the review by Janev {\it et al.} \cite{Janev2003} and the cross-section data for fast-molecule ionization is from the ORNL report by Barnett {\it et al.} \cite{ORNL_5206}.}.
\label{fig:Xsec} 
\end{figure*}

\begin{figure*}[htbp]
\includegraphics[scale=0.7]{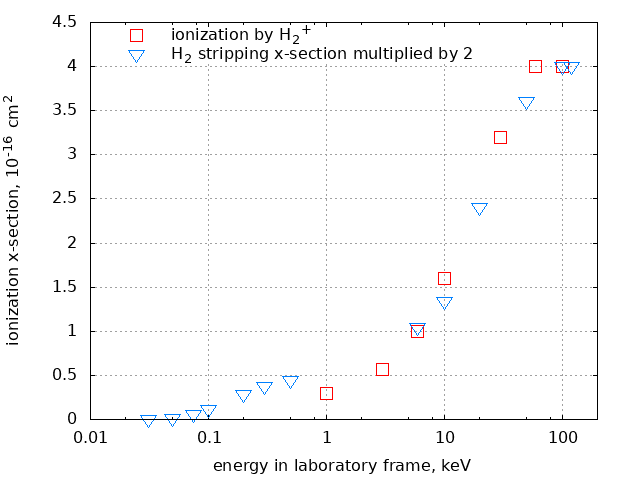}
\caption{Comparison between two sets of reference cross-section data: the cross-section of $\mathrm{H_2}$ impact ionization by $\mathrm{H_2^+}$ ions (squares, Table A.5.26 \cite{ORNL_5206}) is close to twice the stripping cross-section of fast $\mathrm{H_2}$ molecules (triangles, Table A.5.44 \cite{ORNL_5206}).}
\label{fig:consistency} 
\end{figure*}

The adopted cross-sections for gas-phase reactions are plotted in Fig.~\ref{fig:Xsec} vs. kinetic energy in the laboratory frame.
The cross-sections of $\mathrm{H_2}$ impact-ionization by $\mathrm{H_2}^+$ molecular ions of and by fast 
$\mathrm{H_2}$ neutral molecules are taken to be equal, as was done in previous work on the subject \cite{Lauer81}, albeit without express justification. Such approximation is indeed not backed up by theory; however, it proves useful in practice for the $1 -- 100~\mathrm{kV}$ ion energy range of interest to the present study. The close proximity between the two cross-section data sets was first observed by McClure \cite{McClure1964}. The accuracy of the above approximations is illustrated with the aid of Fig.~\ref{fig:consistency}. 
It shows the sets of reference data compiled by ORNL \cite{ORNL_5206} for ion-impact ionization and for electron stripping of fast hydrogen molecules, the latter multiplied by 2 to account for symmetry of the process. The two sets of data, obtained in independent experiments, are indeed in reasonable agreement. The neutral-neutral collision is a symmetric process, and the probability of stripping (electron loss by a fast neutral in the laboratory frame) equals that of producing a slow ion. We utilize the better resolved data set for the stripping cross-section of $\mathrm{H_2^0}$. The relative difference between the two sets of data is higher at lower impact energies, which may be due in part to lower experimental accuracy of measuring the cross-section in this range. This is one of the recognized sources of uncertainty in the resulting prediction and was already encountered in studying the ionization breakdown helium \cite{Liang2017}. To help interpret Fig.~\ref{fig:Xsec}, we note that actual energies of electrons and ions (and consequently, fast neutrals) are determined by the respective free path length. This will be explored in the detailed analysis that will follow. We also note that in a neutral-neutral impact ionization, the fast neutral is lost with a probability of 1/2 (when a fast ion is produced and not a slow one).

\subsection{Surface interactions}
\subsubsection{Secondary electron emission from the cathode}
\begin{figure*}[htbp]
\includegraphics[scale=0.7]{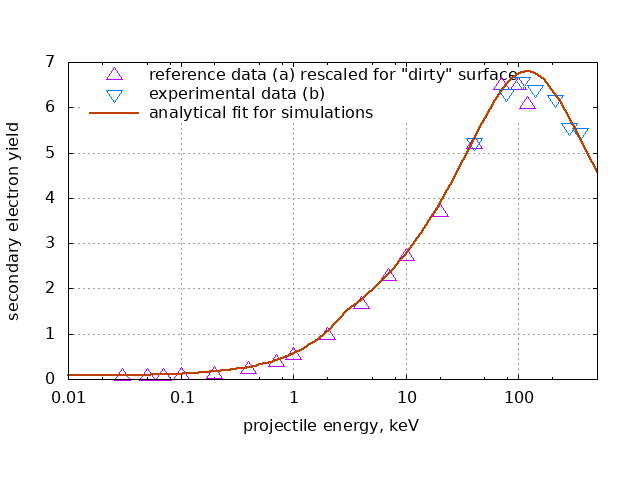}
\caption{Data on electron emission yield of hydrogen molecular ions and fast molecules, with analytical approximation employed in the simulations. Sources : (a) \cite{ORNL_5207} and (b) \cite{MIT1939}. }
\label{fig:IonYield} 
\end{figure*}

Energetic $\mathrm{H_2^+}$ ions and neutral molecules impacting the molybdenum cathode produce high emission yield. As in the past project involving helium, it was important to have reliable data for emission yield from a surface exposed to hydrogen discharge. The approach practiced in previous studies was to rescale the data obtained with a "clean" surface, that is, obtained in high vacuum with a carefully de-gassed sample. Specifically, the ORNL data for molybdenum \cite{ORNL_5207} was rescaled so as to have the maximum yield (at about 100 keV impact energy) equal to $6.5$. We found such approach to be consistent, in fact, with another set of experimental data \cite{MIT1939} obtained specifically in a hydrogen gas environment. Fig.~\ref{fig:IonYield} shows the two sets of data and an analytical fitting curve developed for the present study. 
Note that that there is still a large uncertainty at low energies, as it is not clear whether uniform rescaling is appropriate in that range, where the yield is low. The range of energies in question is still important and past studies have shown \cite{Granzow62} that the choice of approximation even affects the slope of the Paschen curve for applied voltages below 20 kV (which is outside the range specified for the project at hand). The same values of yield are adopted for ions and fast neutrals, because in the case of hydrogen molecules the yield for both is mostly kinetic. The secondary electrons are given a Maxwellian distribution with a temperature of 2 eV. There is no need to specify this distribution more accurately in studying a high-voltage discharge.

\subsubsection{Inelastic electron reflection at the anode}
Electrons emitted from the cathode and also generated in gas-phase ionization undergo mostly ballistic (free-flight) motion towards the anode.
At the anode surface, inelastic reflection process is taking place. For electrons with energies in the range from 10 to several hundred keV, a model for this process is well developed \cite{Darlington75}. The parameters of the process depend on the nuclear charge $Z$ of the material. Since detailed experimental data for molybdenum is not available, we utilized the data for silver ($Z = 47$, vs. the value of 44 for Mo). For normal incidence, the reflection yield agrees with that tabulated in Sec.~4.12 of \cite{ORNL_5207} for molybdenum within the stated accuracy of 10\%. The inelastic reflection model involves specifying the spectrum of the reflected flux for a given energy (not depending on the incident angle when cumulative probability is normalized to 1), and angular dependence of the reflection probability. The spectrum is well approximated by a quadratic function
\begin{equation}
    f(E)=3(E/E_0)^2
    \label{eq:spectrum}
\end{equation}
for the normalized probability density, where $E_0$ is the projectile energy. Monte-Carlo sampling corresponding to Eq.~\ref{eq:spectrum} is performed according to $E/E_0=\sqrt[3]{R}$, where is $R$ is a random number sampled from a uniform distribution on $(0:1]$. Further, the reflection probability dependence on the incidence angle $\theta$ is parametrized as $$\eta(\theta)=0.4\exp\left(0.8(1-\cos\theta)\right).$$ 
Lastly, the inelastically back-scattered flux is emitted isotropically, i.e. according to the cosine law. 

The process of inelastic electron reflection is quite important because it results in a population of electrons trapped by the applied potential in the vicinity of the anode. While the characteristic energies of those electrons are still well above the value at which the electron-impact ionization cross-section has a maximum, the process produces a low-energy tail and the ion flux generated on the anode side undergoes exponential multiplication along the distance towards the anode. 

\subsection{Applying the hydrogen input data to the deuterium isotope}
For deuterium,  the numerical simulations were performed by adapting the kinetic model developed for hydrogen in the following way: The cross-section values for ion-neutral and neutral-neutral ionization in $\mathrm{H_2}$ and $\mathrm{D_2}$ were assumed to be same for the same values of the relative velocity between the target and projectile. The ion-induced and neutral-induced electron emission yields on the molybdenum cathode were assumed identical to those known for $\mathrm{H_2}$ at the same energy, given the absence of reliable experimental and theoretical data.  Available experimental evidence indicates that the yield would be lower for a higher-mass isotope in general, and also specifically for deuterium vs. hydrogen \cite{Krebs1968, Briain1958}

\section{\label{sec:results} Results}
\subsection{Preliminary note on the scaling relation between the Paschen curves for $\mathrm{H_2}$ and $\mathrm{D_2}$}
The basic kinetic model of high-voltage Townsend discharge allows to deduce a simple approximate scaling between the breakdown thresholds in $\mathrm{H_2}$ and $\mathrm{D_2}$ (at least when ions are not fully in a runaway regime). The approximate rule for predicting the breakdown voltage $V_{\mathrm{D_2}}(pd)$ in deuterium when $V_{\mathrm{H_2}}(pd)$ for hydrogen is already known is $V_{\mathrm{D_2}}(pd)\approx 2V_{\mathrm{H_2}}(pd)$. The discussion of the scaling is expanded in the Appendix. Besides offering further insight into the physics of the process, the scaling rule helps to guide the process of numerically predicting the breakdown threshold for $\mathrm{D_2}$ gas. The primary results of this study will now follow below. 

\subsection{\label{subsec:main} Numerically predicted Paschen curves for hydrogen and deuterium}
\begin{figure*}[htbp]
\includegraphics[scale=0.6, angle = 0]{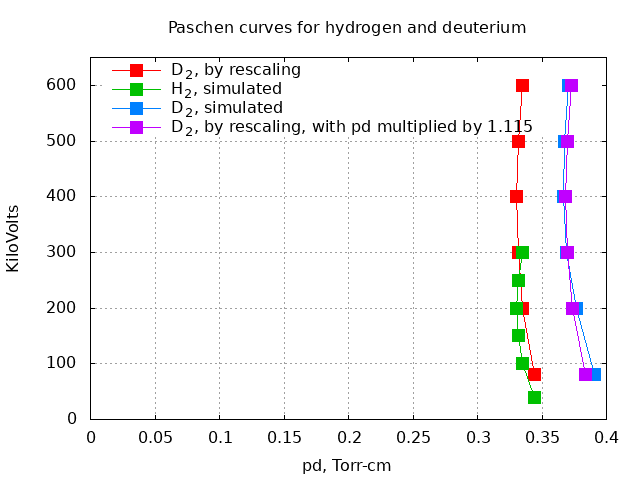}
\caption{Numerically predicted Paschen curves for $\mathrm{H_2}$ and $\mathrm{D_2}$. The $\mathrm{D_2}$ curve predicted by rescaling $(pd,V)\rightarrow(pd, 2V)$ is also plotted.}
\label{fig:Pashen_H2_D2} 
\end{figure*}

The main result of this study is the predicted low-pressure branch of the Paschen curve, 
plotted in Fig.~\ref{fig:Pashen_H2_D2} for both hydrogen and deuterium.
The data points were obtained by the method of steady-state balance discussed in Sec.~\ref{sec:PIC}. The $pd$ value varies within a narrow range, in agreement with previous numerical studies for deuterium \cite{Granzow62} and for hydrogen \cite{Lauer81}. This fact justifies the implicit assumption in our analysis (see Appendix) that the ion kinetics is controlled by the applied voltage and by the energy dependence of the charge-exchange cross-section.
. 

The predicted $\mathrm{D_2}$ Paschen curve is shifted to higher $pd$ values relative to the prediction based on the simple scaling argument. The relative shift is just over 10 percent, indicating that the mass scaling still offers a good insight into the physics of the problem as well as a way to organize and interpret the data. We note again that the actual values of the ion-induced and neutral-induced electron emission yield at the cathode would be lower for $\mathrm{D_2}$ vs. $\mathrm{H_2}$ at a given energy, and the Paschen curve would shift even further towards higher $pd$ values. Therefore, in terms of technological applications, the predicted Paschen curve for deuterium can be viewed as being on the safe side of the uncertainty margin.

\begin{figure*}[htbp]
\includegraphics[scale=0.6]{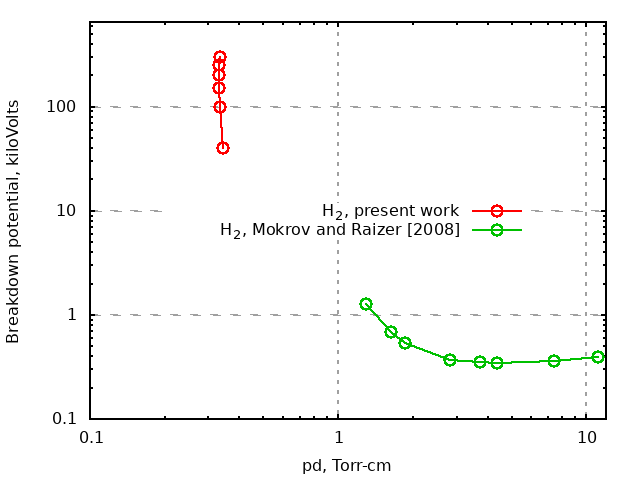}
\caption{The newly obtained Paschen curve data plotted with the data of Mokrov and Raizer \cite{Mokrov2008} for lower voltage.}
\label{fig:Paschen} 
\end{figure*}

To offer an example of how our present results relate to those previously obtained at low voltage, in Fig.~\ref{fig:Paschen} we plot the new data along with those from another numerical study  \cite{Mokrov2008} where results were provided in a tabulated form and the underlying model was described in detail. The latter data is not for molybdenum electrodes but for gold-plated copper; however, the electrode material is of less influence at lower voltage. As noted earlier, the new data is also consistent with prior experimental results \cite{PPPL97} obtained at PPPL; the latter are not available in tabulated form.

A practical result to be noted is that in the range of the applied voltage considered in this study, one can point out a $pd$ value below which the breakdown will not occur. Such value is about 0.33 Torr-cm for ${\mathrm H_2}$ and about 0.36 Torr-cm for ${\mathrm D_2}$.  

\section{\label{sec:PIC} Discussion of numerical simulations}
The specific case below examined in detail to show the numerical work is for the hydrogen gas at the applied voltage of 100 kV. The computational domain is divided into 280 cells ($\Delta x = 50\mu\mathrm m$) which is quite sufficient to resolve all collision lengths. Likewise, the time step is sufficiently small to resolve the collision times.

The charged species are advanced in a constant electric field, while neutrals move inertially. Collisions are processed using a Monte-Carlo procedure, and suitable probabilistic sampling is employed to simulate the electron emission and back-scattering from surfaces (including both the energy spectrum and the angular distribution). 

\subsection{Steady-state ionization balance}
\begin{figure*}[htbp]
\includegraphics[scale=0.7, angle = 0]{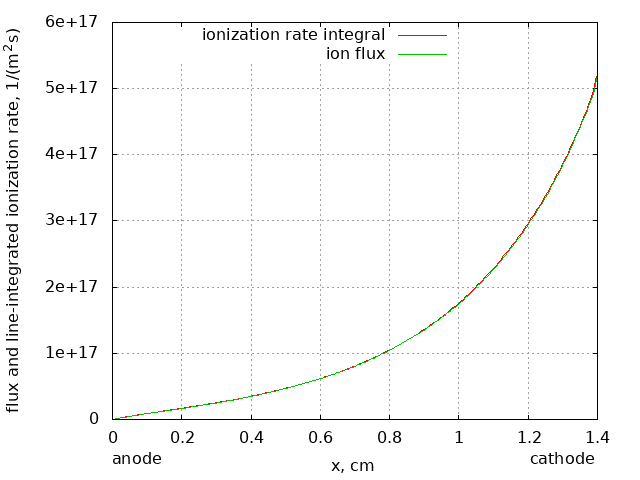}
\caption{Steady-state balance between the ions flux and the running integral of volume ionization rate. The applied voltage is 100 kV.}
\label{fig:Steady} 
\end{figure*}

To identify the breakdown threshold for a given applied voltage, the gas density is varied to achieve a balance between the ionization rate and the flux of ions (the electron flux would then differ from the ion flux by a constant value, representing the electric current). An accurate method to do so, which we deployed in the past \cite{Liang2017}, is to compare the local flux of ions with a running integral of volume ionization rate. This comparison is represented graphically in Fig.~\ref{fig:Steady}. Note that the absolute values on the y-axis are not important because the problem is linear. These values are both proportional to the initial density of the seed plasma in the system. 
We note that the initial slope of the curves near the anode represents the ionization rate $\frac{d\Gamma}{dx}$ due to inelastically back-scattered electrons (also refer to Fig.~\ref{fig:schematic}), where $\Gamma(x)$ is the ion flux with $\Gamma(0)=0$. The detailed kinetics of the particle species and mechanism of the discharge will be examined in what follows.

\subsection{Kinetics of electrons}
\begin{figure*}[htbp]
\includegraphics[scale=0.6, angle = 0]{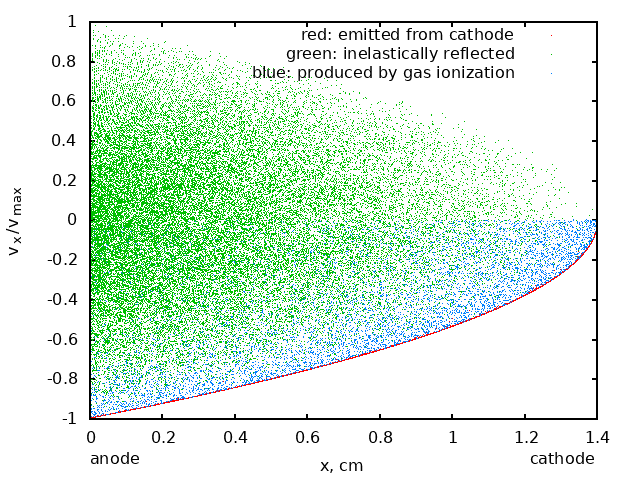}
\caption{Snapshot of electron macroparticles in the phase space for the 100 kV simulation case. The population of trapped inelastically back-scattered electrons near the cathode provides the source of ionization in that region, as shown schematically in Fig.~\ref{fig:schematic}. Specifically, the ionization rate due to back-scattered electrons equals the slope of the curve in Fig.~\ref{fig:Steady} at $x=0$. This back-scattered population is caused by both ballistic (emitted) electrons and those released in gas-phase ionization. The latter is primarily due to ions and fast neutrals.} 
\label{fig:epp} 
\end{figure*}

First, we discuss the kinetic of electrons in a 100~kV discharge with the aid of a phase-space snapshot in a steady state, Fig.~\ref{fig:epp}.
Mapping the particles in the $x--v_x$ plane is a good visualisation of how their velocity distribution varies with the position across the discharge gap, and which groups, identified by their origin, are responsible for forming the distribution.

The $x$-velocity is normalized by its maximal possible value, that is, the one corresponding to the 100~keV kinetic energy.
It is seen that the electrons emitted from the cathode are mostly ballistic (red symbols in Fig.\ref{fig:epp}). Such is also the case for the electrons produced in the gas phase: the energy of an electron originated at $x=x_0$ and observed at $x=x_1$ equals $|eV(x_1-x_0)|/d$ and therefore the energy spectrum at a given point $x_1$ is determined simply by the ionization profile at $x_{1}<x<d$). This is consistent with the volume-produced secondaries (blue symbols in Fig.\ref{fig:epp}) filling the entire phase-space region bounded by the trajectory of an electron emitted from the cathode . The back-scattered population, formed in multiple inelastic reflections, is trapped in the vicinity of the anode. It is this group which is responsible for the ionization rate next to the anode as seen in Fig.~\ref{fig:Steady}.  
\subsection{Kinetics of ions and fast neutrals }
\begin{figure}[htbp]
\includegraphics[scale=0.6, angle = 0]{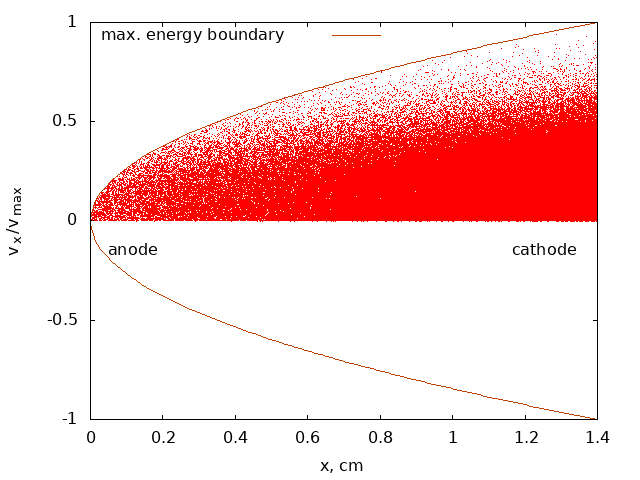}
\caption{Snapshot of ion macroparticles in phase space for the 100 kV simulation case.}
\label{fig:ipp} 
\end{figure}

A phase-space snapshot for ions is shown in Fig.~\ref{fig:ipp}.
Ion velocities are also normalized to the value corresponding to the kinetic energy of 100 keV. Note, however, that there is no ballistic beam, unlike for electrons. This so because the mean free path for charge exchange is less than 1/10th of the discharge gap. Fast neutral molecules are not shown. Their number is of the same order as that of the ions. On the one hand, each ion produces several fast neutrals. On the other hand, in neutral-neutral ionization a fast neutral converts into an ion with a probability of 1/2, ceasing to exist. Only fast neutrals with energies above 100 eV are tracked in the numerical model. At smaller energies, their contribution to gas ionization is negligible. The phase plot shows what is essentially an exponential multiplication of the ion flux towards the cathode. The fluxes of ions and neutrals induce electron emission at the cathode surface; in terms of the ion flux alone the effective yield is about 5.5. 

\begin{figure}[htbp]
\includegraphics[scale=0.6, angle = -0]{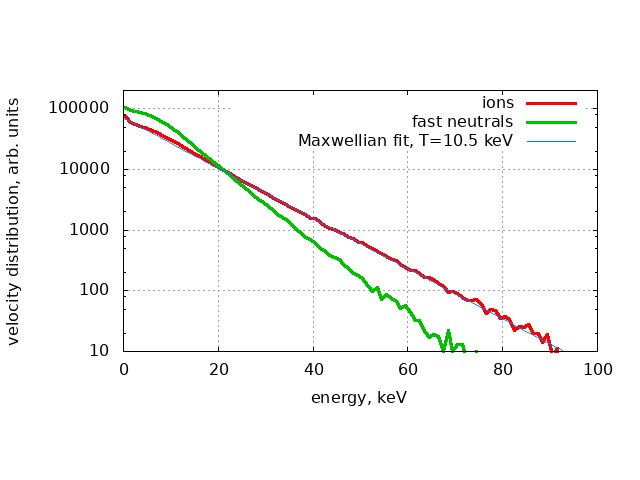}
\caption{Velocity distributions of ions and fast neutrals, plotted vs. energy, for ions and fast neutrals impinging upon the cathode for the 100 kV case.}
\label{fig:VDF} 
\end{figure}

The velocity distributions of ions and fast neutrals at the cathode are shown in Fig.~\ref{fig:VDF}. 
The ion distribution can be approximated with good accuracy by a Maxwellian fit, also shown. This property is connected to the small variation of the charge exchange cross-section vs. energy, as seen in Fig.~\ref{fig:Xsec}. One-sided Maxwellian is a steady-state solution for a velocity distribution of ions in electric field if the charge-exchange cross-section is constant (as seen, for example, in \cite{Lawler1985}). The effective temperature for such Maxwellian distribution is, naturally, a measure of energy gained by the ion over a mean free path distance. 
\subsubsection{Note on ion runaway}
\begin{figure}[htbp]
\includegraphics[scale=0.6, angle = 0]{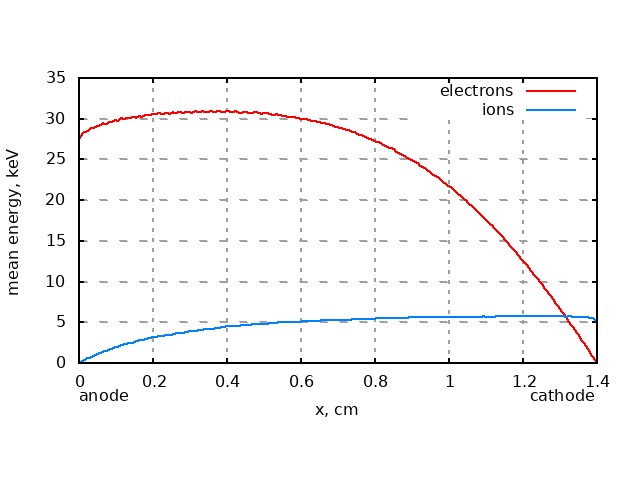}
\caption{Mean energy of ions and electrons vs. distance along the discharge gap for the 100 kV case. All electrons behave non-locally, with energies corresponding to the travel distance. The ion distribution becomes local, but only within 2/3 to 1/2 of the discharge gap towards the cathode. Therefore, kinetic treatment of ions is also required.}
\label{fig:Mean} 
\end{figure}

What is also seen in Fig.~\ref{fig:Xsec} is that the charge-exchange cross-section drops off sharply at energies above 10 keV. If we take the effective energy to be 1/2 of the effective temperature (as is the case for a Maxwellian distribution) then, based on the result for the 100 kV case, an onset of ion runaway would be expected already at applied voltages above 200 kV. Indeed we find that the Paschen curve is sloping to the right between 200 and 300 kV. For the 100 kV case the mean energy of ions vs. $x$, along with that for electrons, is plotted in Fig.~\ref{fig:Mean}. 
It is observed that, although the ion distribution at the cathode is of an equilibrium type, it takes a sizable portion of the inter-electrode distance for this distribution to be established. Indeed, the respective length is estimated by Lawler {\it et al.} \cite{Lawler1985} to be five to six times the free path. Therefore, a kinetic description of the ions is still required. The mean energy of ions impacting the cathode, just above 5~keV, is consistent with the observed effective temperature of 10.5~keV.

\section{\label{sec:conclusion} Conclusions and future work}
The Paschen curves predicted in this study should be of value in the process of designing plasma-based switching devices. An important prediction is the steepness of the curve, implying that a single value of $pd$ chosen with a sufficient safety margin should meet the voltage standoff design specification in the entire range of applied voltage specified for a given application.
The predicted respective $pd$ values are $0.33~\mathrm{Torr-cm}$ for ${\mathrm H_2}$ and about $0.36~\mathrm{Torr-cm}$ for ${\mathrm D_2}$. In applications, these values need to be satisfied with a certain a margin of safety chosen for the design.

The reliability of the model hinges upon the existing uncertainty in the basic input data. The most significant sources of such uncertainty are deemed to be the emission yield and ionization by heavy species at low energies. As noted, there is also a long standing controversy \cite{JETP57,McClure59,JETP62} in experimental literature on whether the $pd$ scaling law applies specifically in the case of hydrogen. For any specific application, additional experiments would be required within the appropriate design space. A more detailed kinetic model would be needed to investigate this question theoretically. 

\begin{acknowledgments}
The information, data, or work presented herein was funded in part by the Advanced Research Projects Agency-Energy (ARPA-E), U.S. Department of Energy, under Award Number DE-AR0001107. The views and opinions of authors expressed herein do not necessarily state or reflect those of the United States Government or any agency thereof.
\end{acknowledgments}

\appendix*
 \section{Approximate scaling to connect the breakdown thresholds in $\mathrm{H_2}$ and $\mathrm{D_2}$ at high voltage}
 
 The scaling argument is based on the following statements:

(1) For ions as well as fast neutral molecules, the impact-ionization cross-section is the same for $\mathrm{H}_2$ and $\mathrm{D}_2$ isotopes with the same velocity. The same applies to the energy-dependent electron emission yield.

(2) Ions have a quasi-equilibrium distribution with a characteristic energy equal to the product of the applied electric field by the mean free path for charge exchange. The fast neutrals produced in charge exchange will have the same characteristic energy.

(3) Ionization responsible for generating the initial ion flux in the vicinity of the anode is due to electrons, from both ballistic (primary flux) and reflected distributions present in that area. For both populations, the mean energy scales proportionally to the applied potential. It is seen that this scaling does not actually involve the ion mass unlike the energy of the heavy species. However, in the present model the electron-impact ionization near the anode plays a role quite similar to that of the cathode emission yield in the standard Townsend theory. Accordingly, its dependence on the applied voltage $V$ is off less significance than the ionization coefficient which determines exponential multiplication of the ion flux. It will be shown that this dependence is in inverse proportion to $V$; therefore the re-scaled Paschen curve will underestimate the $pd$ value for the given value of the voltage. This is indeed observed in the numerical results.

First, let us suppose that electron-impact ionization near the anode is simply proportional to the primary flux. This ignores energy dependence of the ionization cross-section. Then the balance equation determining the breakdown threshold will involve functions that depend only on $pd$ and  characteristic ion velocity $\sqrt(\frac{V}{Mn\sigma_{cx}})$. The latter determines ionization cross-sections and therefore the eigenvalue (exponential index) of the problem will scale as $pd F(\frac{V}{Mn\sigma_{cx}})$, where $F$ is a function with the same dimension as cross-section. Likewise, the flux-averaged electron emission yield at the cathode will have a form $G(\frac{V}{Mn\sigma_{cx}})$. It follows that a given point $(pd, V)$ on the Paschen curve for hydrogen will map into a point $(pd, 2V)$ on the Paschen curve for deuterium. 

\begin{figure}[htbp]
\includegraphics[scale=0.6]{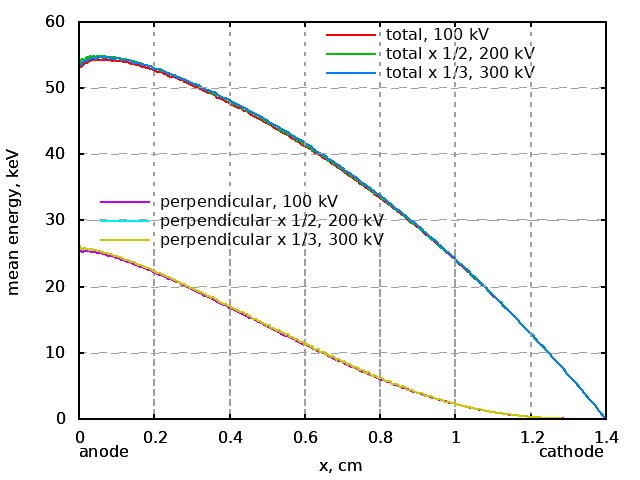}
\caption{Mean energy of electrons vs. position in the discharge gap, }
\label{fig:energy_scaling} 
\end{figure}

\begin{figure}[htbp]
\includegraphics[scale=0.6]{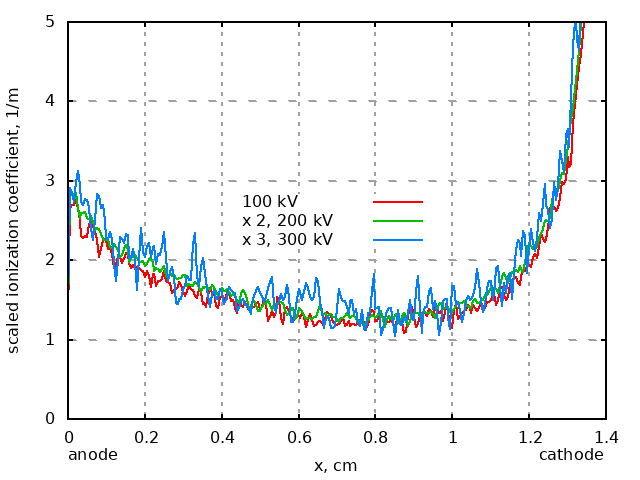}
\caption{Scaled coefficients of electron-induced ionization coefficients for three different values of applied voltage}
\label{fig:Ze_scaling} 
\end{figure}

A brief discussion is needed of the electron-induced ionization, to show that it scales inversely with applied voltage. Fundamentally, this is so because (a) the electron-impact ionization cross-section falls of as $1/\epsilon$ (where $\epsilon$ is electron energy) for $\epsilon$ much higher than the ionization potential of the $\mathrm{H}_2$ or $\mathrm{D}_2$ molecule, and (b) the mean energy of electron population at a given position within the discharge gap scales proportionally to the applied voltage. To illustrate the latter point, we refer to Fig.~\ref{fig:energy_scaling}.
The plots show mean energies corresponding to velocity components parallel and perpendicular to the applied electric field, for three cases with the applied voltage of 100, 200, and 300 KV. The respective curves coincide when the energy is divided by voltage. We note that the transverse motion occurs in the trapped back-scattered population present near the anode. Therefore, the scaling, as expected, applies to both ballistic and back-scattered electrons. 
Next, to aid the discussion, we turn to Fig.~\ref{fig:Ze_scaling}.    
The plots show local electron-induced ionization coefficients for the three cases considered above (100, 200, 300~kV), multiplied, respectively by the factors of 1, 2, and 3. The scaled curves coincide, indicating that the volume ionization rate indeed scales in inverse proportion to the applied voltage. This concludes our qualitative discussion of the Paschen curve scaling. 

We note that the $V/M$ breakdown voltage scaling for hydrogen and deuterium has been reported before \cite{Miley1997}, in connection with Monte-Carlo modeling of electrostatic confinement fusion (ECF) experiments.
The Paschen curves obtained for hydrogen and deuterium were found to coincide if $V/M$, where $M$ is mass in atomic units, was used as the coordinate instead of $V$ for the breakdown potential. 
The calculations by Miley {\it et al.} were performed for a spherical configuration with a highly transparent spherical-grid anode at the center that allows accelerated ions to fly through. This explains lower $pd$ values than those for planar discharge with solid electrodes (In ref.\cite{Miley1997}, the standard Paschen curve is also shown for reference in another plot).
 

\bibliography{Paschen_H2_D2}

\end{document}